\documentclass[12pt,tightenlines,eqsecnum,floats,shownopacs,nofootinbib,amsmath,amssymb,aps,prd]{revtex4}

\usepackage{amssymb}
\usepackage{amsmath,amssymb,amsfonts,amsthm}
\usepackage{graphicx}

\usepackage{setspace}
\usepackage[colorinlistoftodos]{todonotes}
\usepackage{epstopdf}

\def\be{\nopagebreak[3]\begin{equation}}
\def\ee{\end{equation}}
\def\ba{\nopagebreak[3]\begin{eqnarray}}
\def\ea{\end{eqnarray}}

\def\lp{\ell_{\rm Pl}}

\def\f{\frac}

\def\B{{\rm B}}

\def\t{\tilde}
\def\h{\hat}

\def\x{\vec{x}}

\def\R{\mathcal{R}}

\def\U{\mathfrak{A}}

\def\T{\mathcal{T}}

\def\R{\mathcal{R}}
\def\Q{\mathcal{Q}}

\begin{document}

\title{Symmetry Reduced Loop Quantum Gravity:\\ A Bird's Eye View}
\author{Abhay Ashtekar}
\email{ashtekar@gravity.psu.edu}
\affiliation{Institute for Gravitation and the Cosmos \& Physics
  Department, Penn State, University Park, PA 16802, U.S.A.}

\begin{abstract}

This is a brief overview of the current status of symmetry reduced models in Loop Quantum Gravity (LQG), focussing on the very early universe. Over the last 3 years or so the subject has matured sufficiently to make direct contact with observations of the Cosmic Microwave Background (CMB). In particular, thanks to an unforeseen interplay between the ultraviolet features of quantum geometry and the infrared properties of quantum fields representing cosmological perturbations, \emph{Planck scale effects} of LQG can leave imprints on CMB \emph{at the largest angular scales}. In addition to a summary of these results, the article also contains a critical discussion of the symmetry reduction procedure used in discussions of quantum cosmology (and quantum black holes). 


\end{abstract}


\maketitle

\section{Symmetry reduction}
\label{s1}

It is a common strategy in physics to probe consequences of a theory by truncating it to a sector that is tailored the specific issues one wishes to explore. For example, in the study of gravitational waves, one truncates general relativity (GR) to its asymptotically flat sector and analyzes the structure of solutions at null infinity. Similarly, in quantum electrodynamics one truncates the full theory to the sector with a specified number of incoming and outgoing photons and electrons and then analyzes contributions to the scattering amplitude by various permissible processes in the interaction region. 

For certain physical problems it is natural to \emph{tailor the truncation to symmetries}. For example, to understand thermodynamic properties of black holes in equilibrium, one restricts oneself to (asymptotically flat) \emph{stationary} solutions of GR admitting event horizons and studies processes cause a transition from one stationary black hole to a nearby one. Similarly, to analyze astrophysics around black holes, one focuses on the sector of GR consisting of stationary black holes \emph{together with linear perturbations} on these backgrounds. Another example is provided by the investigations of black hole formation through gravitational collapse and subsequent evaporation due to quantum processes. Here, one typically truncates the theory to its \emph{spherically symmetric}, dynamical sector because of the general expectation that this sector already captures the essential conceptual issues.

One of the most successful application of this symmetry reduction strategy in all of physics is to the cosmology of the early universe. Here, one focuses on homogeneous, isotropic --i.e., Friedman, Lema\^itre, Robertson, Walker (FLRW)-- geometries of GR and \emph{linear} perturbations on them. In the passage to quantum theory, one keeps the FLRW geometry classical and quantizes only the perturbations. In particular, the inflationary scenario is developed within this truncation scheme. At first the strategy appears to be a drastic oversimplification from the perspective of the full theory. Indeed, a priori one would have expected the early universe to be extremely complicated since, if we were to move back in time, matter in the universe would seem to clump more and more, making space-time geometry in the early epoch very complicated due to non-linearities of GR. But observational missions --particularly WMAP and PLANCK-- have shown that the theoretical strategy is extremely effective in capturing the physics of the very early universe: the cosmic microwave background is astonishingly isotropic and the universe is much simpler than one would have a priori expected. What first seems like a naive truncation of a highly non-linear theory in fact successfully encodes the essence of physics governing the early universe.
 
In contemporary \emph{quantum} cosmology, one uses the same strategy to further extend our theoretical reach from the inflationary scale to the Planck era, i.e., over some 12 orders of magnitude of density and curvature. One now begins with \emph{quantum} FLRW geometries. Since the 1970s, it has been hoped that this truncation to the homogeneous isotropic models would be sufficient to capture the salient quantum effects that are expected to tame cosmological singularities. The voluminous work on Loop Quantum Cosmology (LQG) carried out over a decade, starting in the early 2000s, has shown that this strategy does indeed lead to a natural resolution of the big bang singularity. (For reviews, see, e.g., \cite{asrev,ps3}.) The essential reason can be traced back to to specific quantum geometry effects that are hallmarks of full Loop Quantum Gravity (LQG). The focus over the last five years or so has shifted to the study of quantum fields representing cosmological perturbations that now propagate on \emph{quantum FLRW geometries}. This body of ongoing work has revealed unforeseen effects of pre-inflationary dynamics, especially in the Planck regime. In particular, there is an interesting interplay between the quintessential features of quantum geometry in the deep \emph{ultraviolet} that cure the big bang singularity and the \emph{infrared} properties of cosmological perturbations that leave an imprint on CMB at the largest angular scales (for a summary, see \cite{ab,bcgm}). Thus, LQC is now sufficiently mature to make direct contact with observations of the very early universe. On the mathematical side, one can trace back several advances to the development of quantum field theory on cosmological \emph{quantum} geometries \cite{akl,aan2}. Subsequently, this procedure was adopted to problems involving black holes through quantum field theory on \emph{quantum} black hole backgrounds \cite{gp1,gp2,gop}.

In section \ref{s2}, I will sketch the results on singularity resolution. This discussion will be rather brief since these results are older and there are already several detailed review articles that summarize the situation. However, I will present the material from a fresh perspective. Further recent developments in this area are summarized in the articles by Craig, Corichi and Singh in this special issue. In section \ref{s3}, I will discuss in much greater detail the current status of the ongoing analysis cosmological perturbations starting from the deep Planck regime. As Chapters by Agullo, Barrau, Mena-Marugan and Wilson-Ewing in this issue illustrate, different groups are currently pursuing closely related but technically distinct directions. I will not be able cover all of them in this brief overview. Rather, I will focus on ideas that have led to the most comprehensive and detailed observational predictions and, at the same time, introduced novel ideas to address conceptually important issues such as the selection of initial conditions. Taken together, recent advances made by various groups illustrate how much LQG has evolved since the 1990s. Since the focus then was on creating foundations for a non-perturbative and background independent approach, most of the work was in the realm of mathematical physics, rather far removed from observations. Now the focus is on the a interplay between theory and observation.

In light of these advances, it is natural to return to the general issue of symmetry reduction and examine the underlying strategy more carefully. In quantum cosmology, one first carries out symmetry reduction of classical GR and then constructs quantum theory of just this sector. While the strategy seemed `obvious' as the starting point in the 1970s, now that quantum cosmology has progressed sufficiently to make contact with observations, it is appropriate to evaluate it critically. Is the overall procedure sufficiently reliable for us to trust the predictions on dynamics of the few observables that are relevant for the large scale structure of the early universe? Or, would a full quantum gravity theory such as LQG lead to qualitative changes in this dynamics? Since the strategy underlies \emph{all} approaches to quantum cosmology, these questions are quite general and not specific to LQC. Within LQC, they apply to all directions that are being currently pursued to  probe pre-inflationary dynamics. Section \ref{s4} discusses this issue and also includes comments on the investigations of quantum black holes within LQG.

\section{Singularity Resolution due to Quantum Geometry}
\label{s2}

Every expanding FLRW solution of GR, has a big bang singularity if matter satisfies the standard energy conditions. This is a strong curvature singularity at which physics just comes to a halt in classical GR. However, as Einstein \cite{ae1} himself emphasized already in 1945,
\begin{quote}{\emph{``One may not assume the validity of field equations at very high density of field and matter and one may not conclude that the beginning of expansion would be a singularity in the mathematical sense.''}} \end{quote}
By now we know that validity of classical physics cannot be taken for granted in the astronomical world because quantum phenomena can have dramatic effects even on large, macroscopic systems. For example, the Fermi degeneracy pressure --a quintessentially quantum effect-- is responsible for the very existence of neutron stars. At nuclear densities of about $10^{15} {\rm gm/cm^{3}}$, it becomes enormous and can counterbalance the mighty gravitational attraction, thereby halting the gravitational collapse. Where classical physics would have predicted the formation of a singularity, quantum mechanics intervenes and leads to new equilibrium configurations.

Since the 1970s, it has been a cherished hope that quantum gravity effects would similarly step-in and resolve the big-bang singularity in cosmology. Thus, singularities of general relativity can be regarded as outstanding opportunities to probe physics beyond Einstein. Since gravity is encoded in geometry in GR, it is natural to expect that a quantum theory of gravity will require or lead to an appropriate quantum generalization of Riemannian geometry. LQG has taken this premise seriously and systematically constructed a specific theory of quantum  Riemannian geometry rigorously (for reviews, see, e.g., \cite{alrev,crbook,ttbook}). This theory brings out a fundamental discreteness at the Planck scale. In particular, one can naturally define self-adjoint operators representing geometric observables --such as areas of physical surfaces and volumes of physical regions \cite{al5,al-vol}. It has been shown that geometry is quantized in the direct sense that eigenvalues of geometric operators are discrete \cite{rs,al5,al-vol,length}. In particular, there is a smallest non-zero eigenvalue  of the area operator, $\Delta_{o}\,\lp^{2}$. Thus the theory provides a new fundamental dimensionless parameter $\Delta_{o}$, called the \emph{area gap}. This \emph{microscopic} parameter sets the scale for new phenomena.%
\footnote{In the early LQG and LQC literature, one started with classical GR and passed to connection variables using a canonical transformation that involves the mathematical, Barbero-Immirzi parameter $\gamma$. The area gap $\Delta_{o}$ was then expressed in terms of $G, \hbar$ and $\gamma$. However, as pointed out in \cite{ag1}, it is the area gap that is the fundamental \emph{physical} parameter of the theory from the perspective of the final quantum geometry, and potential observational consequences. One can easily eliminate the `mathematical parameter' $\gamma$ in favor of the `operational parameter' $\Delta_{o}$ simply by setting $\gamma:= \Delta_{o}/(4\sqrt{3}\,\pi)$ in all expressions. Furthermore, this shift makes the classical limit more transparent and unambiguous in LQC and clarifies conceptual issues in the investigations of black hole entropy.}

 \begin{figure}
\begin{center}
\includegraphics[width=4.7in,height=2.7in]{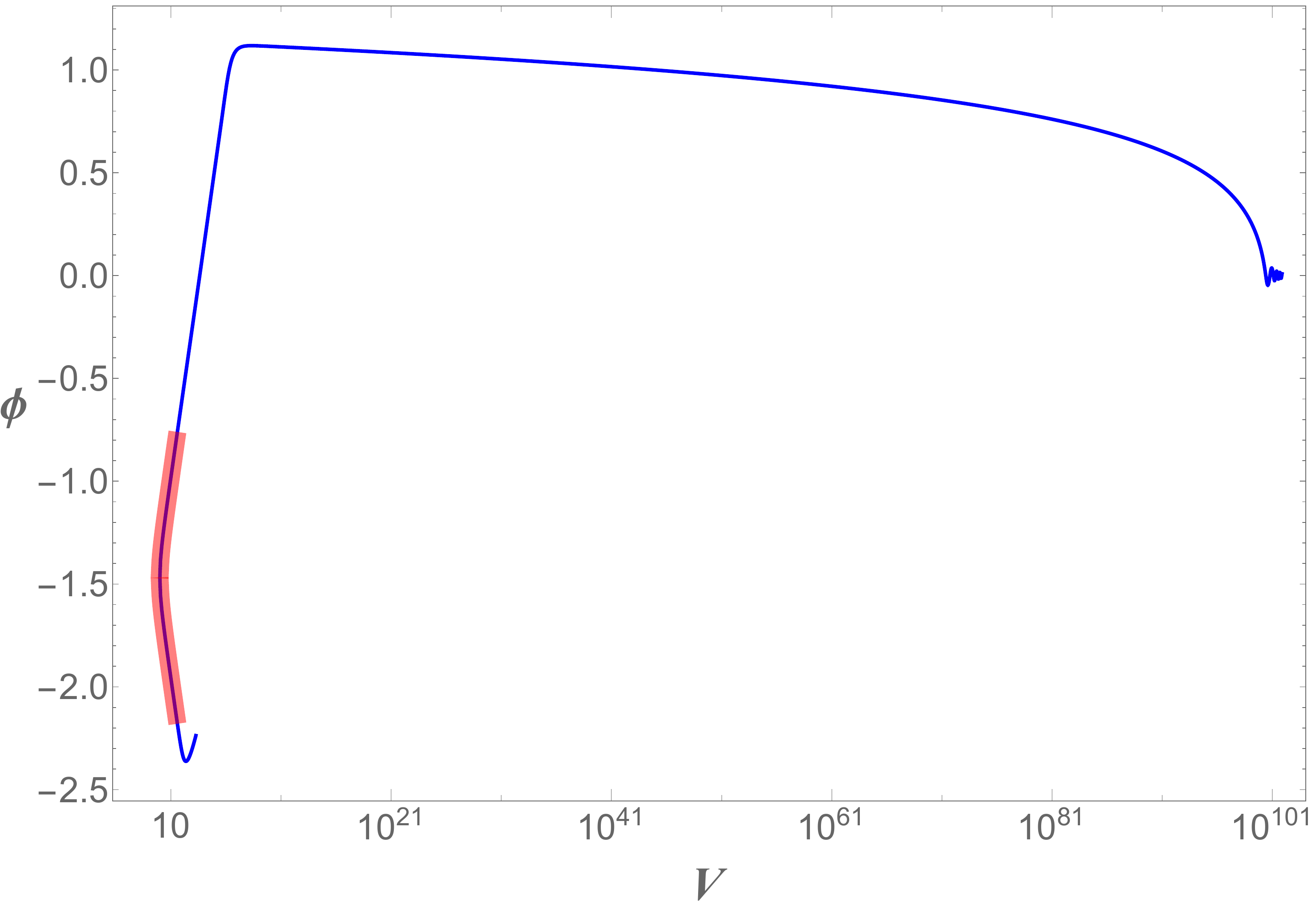}
\caption{\label{fig2} {\footnotesize{An effective LQC trajectory in presence of an inflation with the Starobinsky potential, $V(\phi) = (3M^{2}/32\pi)\,(1-\exp -\sqrt{(16\pi/3)\phi})^{2}$, analyzed in \cite{bg1,bg2}. Here $V \sim a^3$ is the volume of a fixed fiducial region. The long (blue) sloping line at the top depicts slow roll inflation. As $V$ decreases (from right to left), we go back in time and the inflaton $\phi$ first climbs up the potential, then turns around and starts going descending. In classical general relativity, volume would continue to decrease until it becomes zero, signaling the big bang singularity. In LQC, the trajectory bounces and volume never reaches zero; the entire evolution is non-singular. Situation is similar with other potentials. I use the Starobinsky potential as a concrete example because it is phenomenologically favored by the Planck data.}}}
\end{center}
\end{figure}

An analogy with the theory of superconductivity is instructive to bring out this fact. Recall that in the theory of superconductivity, the energy-gap $\Delta_{\rm E}$ --the energy needed to break loose the electrons in a cooper pair-- serves as the microscopic parameter and determines the values of macroscopic parameters such as the critical temperature at which superconductivity sets in: $T_{\rm crit} = ({\rm const})\,\, \Delta_{\rm E}$, where the constant depends on the material. In LQC, it is again the structure of the Hamiltonian that determines the key microscopic parameter of the quantum theory. Recall that in classical GR the Friedmann equation serves as the Hamiltonian constraint that generates the FLRW phase-space dynamics. In LQC, the quantum Friedmann equation turns out to be a \emph{difference} equation, where the step-size is dictated by $\Delta_{o}$. In the limit $\Delta_{o} \to 0$ the step size goes to zero and the LQC Hamiltonian constraint tends to the Wheeler-DeWitt differential equation \cite{aps2,aps3,acs}. The reason why the area gap enters is that in LQC the operator corresponding to the classical curvature tensor is non-local, given by the holonomy of the gravitational connection around suitable loops that enclose the minimum possible area.  Detailed calculations show that the matter density operator $\h\rho$ is bounded above by a universal constant $\rho_{\rm sup}$, and the value of this new macroscopic parameter  is determined by the microscopic parameter $\Delta_{o}$ \cite{aps3,acs,ag1}: 
\footnote{The numerical factor $0.41$ refers to the commonly used value of the area gap that arises from black hole entropy considerations. It could be slightly modified by more sophisticated considerations. For comparisons with observations, this would alter the number of pre-inflationary e-folds (between the LQC bounce and the onset of the slow roll phase) but not affect any of the main conclusions.} 
\be \rho_{\rm sup}\, =\, \frac{{\rm 18\pi}}{G^{2}\hbar\, \Delta^{3}_{o}}\,\, \approx \,\,0.41 \rho_{\rm Pl}.\ee
Note that if we let the area-gap $\Delta_{o}$ to go to zero  --i.e., ignore the quantum nature of geometry underlying LQG--  $\rho_{\rm sup}$ diverges, quantum geometry effects of LQC disappear, and we are led back to the big bang of GR. This is analogous to the fact that, as the the energy gap $\Delta_{\rm E}$ goes to zero, the critical temperature goes to zero and we no longer have the novel phenomenon of superconductivity.

It is important to note that the LQC singularity resolution does not occur simply because the difference equation enables one to  `jump over' the zero volume. Rather, the zero volume state simply decouples from the rest. More precisely, the singularity is resolved in the following sharp sense:  physical observables, such as energy density and curvature which diverge at the big bang in GR, have a \emph{finite} upper bound on the entire physical Hilbert space of states $\Psi_{o}$ of LQC.%
\footnote{Sometimes apparently weaker notions of singularity
resolution are discussed. Consider two examples \cite{kks}. One may
be able to show that the wave function vanishes at points of the
classically singular regions of the configuration space. However, if
the \emph{physical} inner product is non-local in this configuration
space ---as the group averaging procedure often implies--- such a
behavior of the wave function would not imply that the probability
of finding the universe at these configurations is zero. The second
example is that the wave function may become highly non-classical.
This by itself would not mean that the singularity is avoided unless
one can show that the expectation values of a suitable family of Dirac
observables which become classically singular remain finite there.}
(This resolution has also been understood in detail in the sum over histories approach \cite{ach3,chn} and the `consistent histories' framework \cite{consistent}, described in the article by Craig in this issue.) In every physical state, the expectation value of matter density achieves a maximum value $\rho_{\rm max} < \rho_{\rm sup}$, at which the universe bounces, avoiding the formation of a singularity \cite{aps3,acs}. States $\Psi_{o}$ for which $\rho_{\rm max} \approx \rho_{\rm sup}$ are sharply peaked on trajectories satisfying certain \emph{effective}, quantum corrected equations. In particular, the Friedmann and Raychaudhuri equations of GR 
\be H^{2} \equiv (\dot{a}/a)^{2} = 8\pi G \rho; \quad{\rm and}\quad \dot H = -4 \, \pi G\, (\rho\,+p)\, \ee 
are replaced by the effective equations that incorporate leading order quantum corrections:
\be H^{2} = 8\pi G \rho\,\, \big(1 -\f{\rho}{\rho_{\rm sup}}\big)\, \quad{\rm and} \quad
\dot H =  -4 \, \pi G\, (\rho\,+p)\, \Big(1- 2\f{\rho}{\rho_{\rm sup}}\Big).\,  \label{effeq}\ee
Note that the GR equations are excellent approximations to the quantum corrected ones so long as the matter density $\rho$ is much smaller than $\rho_{\rm sup}$ which is of the order $\rho_{\rm Pl}$. To understand the singularity resolution, let us focus on the quantum corrected Friedmann equation. At the bounce the right side vanishes and $\dot{a}$ changes sign. While the effective trajectories are well-approximated by the classical FLRW solutions for energy densities, say, $\rho \lesssim 10^{-3}\rho_{\sup}$, in the Planck regime there are significant departures from GR. In effect, quantum geometry introduces \emph{a novel repulsive force} which is completely negligible once the curvature is below $\sim 10^{-3}$ in Planck units but which rises very rapidly above this scale and becomes so strong as to overwhelm  the classical attraction and cause the universe to bounce. In the spatially flat, k=0 FLRW models, as one goes back in time from the bounce, the universe expands again and dynamics quickly becomes well approximated by Einstein's equations. FIG. \ref{fig2} shows a bouncing effective trajectory for the the k=0 FLRW universe sourced by an inflaton $\phi$ in the phenomenologically favored Starobinsky potential. But qualitatively the behavior is the same for other regular potentials. All solutions to the set of effective Friedmann and Raychaudhuri equations are everywhere regular and singularity free.

While singularity resolution in the quantum theory holds for all states in the LQC physical Hilbert space, the effective equations (\ref{effeq}) were derived starting from quantum states that are sharply peaked Gaussians \emph{at late times}, and evolving them back in time. The full quantum evolution of these states $\Psi_{o}$ exhibits a bounce at a matter density that is (smaller than but) so close to $\rho_{\rm sup}$ that it could not be distinguished from $\rho_{\rm sup}$ even in accurate numerical simulations. It was found that these states remain sharply peaked throughout the evolution and the peak follows the effective trajectories given by (\ref{effeq}). More recently, dynamics of states that are \emph{not} sharply peaked at late times was also investigated for completeness \cite{dgs2,dgms,ag1}. As one would expect from the fact that $\rho_{\rm sup}$ is the upper bound of the spectrum of the density operator $\h{\rho}$, now the bounce occurs at densities $\rho_{\rm B} < \rho_{\rm sup}$ \cite{cm1}. Furthermore, the value of $\rho_{\rm B}$ is dictated by the relative dispersion in the volume operator at late times; larger the dispersion, lower the value of $\rho_{\rm B}$:
\be \rho_{\rm B} \approx \Big(1+\big({\Delta \h{V}}/{\langle \h{V}\rangle}\big)^2_{\rm late\,\,\, time}\Big)^{-1}\,\rho_{\rm sup}. \ee
Finally, it turns out that the expectation values of the volume (or the scale factor) operator again follow certain trajectories that can be derived starting from a generalized effective Hamiltonian. Surprisingly, the resulting Friedmann and Raychaudhuri equations are given by a straightforward generalization of (\ref{effeq}): the only change is that $\rho_{\rm sup}$ is replaced by $\rho_{\rm B}$! Thus, again, classical GR becomes a good approximation for these widely spread states once the matter density falls well below $\rho_{\rm B}$. I should emphasize that if one were to take such states as serious candidates to describe the early universe, one would need a new mechanism --such as the one described in \cite{stamp}-- to drastically reduce the dispersions in the FLRW quantum geometry and make them negligible, say by the end of inflation. The current investigations are \emph{not} driven by such scenarios. Rather, the primary motivation is to probe robustness of the qualitative aspects of singularity resolution and of the subsequent evolution of cosmological perturbations. The detailed investigations have confirmed this robustness \cite{ag1,dgs2,dgms}.
\\

So far, I have focused on the k=0, $\Lambda=0$ FLRW models. However, by now a large number of cosmological models have been studied in detail in LQC, including the closed and open FLRW models, models with a cosmological constant, the Bianchi models and the Gowdy models which incorporate the simplest types of inhomogeneities in full GR (see for example, \cite{apsv, warsaw1,bp,pa,kp1,awe2,madrid-bianchi,awe3,we,gs,cm2,swe,ps2,hybrid1,hybrid2,hybrid3,hybrid4,hybrid5}.) Detailed investigations were carried out using Hamiltonian methods and canonical quantization, complemented by a sum over histories analysis \`a la Feynman for FLRW and Bianchi I models \cite{ach3,chn}. In all cases, the singularity is resolved. (This is notable already for Bianchi models where, because the anisotropic shear terms grow as $1/a^{6}$ near the big bang in GR, as noted in section \ref{s1}, singularity resolution has been difficult in other approaches.) In these investigations, numerical simulations played an important role. Thanks to contributions from experts in numerical GR, LQC simulations have become quite sophisticated, providing unforeseen results and guiding conjectures for theoretical analysis \cite{gs,cm1,cm2,dgs2,dgms,numerical-rev}. The combination of analytical and numerical investigations has brought out a general pattern that can be qualitatively summarized as follows. GR is an excellent approximation to LQC at low curvature. But when curvature grows to $\sim 10^{-3}$ in Planck units, a novel repulsive force originating in quantum geometry starts to become significant. It has a `diluting effect' that counteracts the continued growth of curvature that would have occurred in GR. As I noted in the beginning of this section, the fact that repulsive forces with origin in quantum mechanics can have macroscopic, even astronomical, implications is already known from the existence of neutron stars. But in LQC, the origin of the repulsive force lies not in the fermionic nature of matter, but rather in specific properties of the LQG quantum geometry. (For a review of singularity resolution in LQC, see, \cite{asrev,ps3}.) 

This completes the brief overview of conceptual issue of singularity resolution due to quantum geometry effects of LQG. In the next section, I will discuss more recent developments on predictions of LQC that have created a new interface between theoretical investigations of Planck scale geometry and observations of the very early universe.


\section{Cosmological perturbations and observations}
\label{s3}

Perhaps the most significant reason behind the rapid and spectacular
success of quantum mechanics, especially in its early stage, is the
fact that there was already a significant accumulation of relevant
experimental data, and further experiments to weed out ideas could
be performed on an ongoing basis. Unfortunately this has \emph{not} been the case for relativistic gravity simply because theory has raced far ahead of technology. Indeed, even in the classical regime, the first test of general relativity in the strong field regime was announced only in February of 2016 by the LIGO team!

However, fortunately, the spectacular observations of the very early universe over the last two decades are now unveiling a window into the potential quantum gravity effects in the very early universe.  Perhaps the most remarkable aspect of these developments is that, through a confluence of theory and observations, \emph{we can now trace-back the origin of the large scale structure of the universe to the vacuum fluctuations of quantum fields in the very early universe.} These quantum fields represent the scalar and tensor perturbations on a classical FLRW space-time. Thus we already have good evidence for the quantum nature of scalar fluctuations, i.e., for perturbative quantum gravity. In this section I will discuss how this interplay between theory and observations could also inform us about the Planck regime of full quantum gravity. 

Since a majority of quantum gravity researchers focus primarily on mathematical physics issues, they are typically not familiar with observational issues in cosmology. Therefore, in section \ref{s3.1}, I will briefly outline the inflationary paradigm. In section \ref{s3.2}, I will explain why, contrary to a general impression among cosmologists, pre-inflationary dynamics does matter, and how Planck scale physics could in principle leave imprints on the largest angular scales of CMB observations. In section \ref{s3.3} I will sketch some of the recent advances.

\subsection{Inflationary Scenario: Successes and Open Issues}
\label{s3.1}

Almost every well-developed cosmological scenario assumes that the early universe is well described by a FLRW solution to Einstein's equations with suitable matter, \emph{together with} first order perturbations. As I already mentioned, the background space-time is treated classically, as in GR, and the perturbations are described by \emph{quantum fields}. Thus, cosmological perturbation theory in GR and quantum field theory (QFT) on FLRW space-times provide the platform for this analysis. It is fair to say that among the mainstream scenarios, inflationary paradigm has emerged as the leading candidate. In addition to the common assumption described above, this scenario posits:

\begin{itemize}

\item Sometime in its early history, the universe underwent a
    phase of near-exponential expansion. This was driven by the slow roll of a scalar field in a suitable potential causing the Hubble parameter to be nearly constant.

\item Fourier modes of the quantum fields representing
    perturbations were initially in a specific state, called the
    Bunch-Davies (BD) vacuum, for a certain set of co-moving
    wave numbers $(k_o, 2500k_o)$ where the physical wave length
    of the mode $k_o$ equals the radius $R_{\rm LS}$ of the
    observable universe at the surface of last scattering.%
\footnote{Strictly speaking, the BD vacuum refers to de
Sitter space; it is the unique `regular' state which is
invariant under the full de Sitter isometry group. During
slow roll, the background FLRW geometry is only
approximately de Sitter whence there is some ambiguity in
what one means by the BD vacuum. One typically assumes that
all the relevant modes are in the BD state a few e-foldings 
before the (longest wavelength, observable) mode $k_o$ leaves 
the Hubble horizon. Throughout this part, by BD vacuum I mean this
state.}

\item Soon after any mode exits the Hubble radius, its quantum
    fluctuation can be regarded as a classical perturbation and
    evolved via linearized Einstein's equations.
\end{itemize}

One then evolves perturbations starting from the onset of the slow
roll till the end of inflation using QFT on FLRW space-times
and calculates the power spectrum (see, e.g.,
\cite{ll-book,sd-book,vm-book,sw-book,gr-book}). The power spectrum provides us with a Gaussian peaked at the zero mean value. The mean value and the standard deviation of the Gaussian provides a probability distribution function on the \emph{classical} phase space of linearized fields. This distribution is then evolved using standard numerical codes that have built into them techniques to incorporate various astrophysical effects. The result is a spectrum of fluctuations at the surface of last scattering, which occurred at approximately 380,000 years after the big bounce. \emph{This prediction is in excellent agreement with the inhomogeneities observed in the CMB.} 

These inhomogeneities, in turn, serve as seeds for structure formation as slightly over dense regions become denser due to gravitational attraction, sucking away matter from the less dense regions which then become even more sparse. Supercomputer simulations of this subsequent non-linear evolution have shown that the result is excellent qualitative agreement with the large scale structure of the universe we observe today. This is the precise sense in which \emph{the origin of the qualitative features of the observed large scale structure can be traced back to the fluctuations in the quantum vacuum at the onset of inflation.} This is both impressive and intriguing.

Nonetheless, over the years, the inflationary paradigm has been criticized especially in the relativity community on grounds that are most eloquently expressed by Roger Penrose (see, e.g., \cite{rp1}). However, these criticisms refer to the motivations that were originally
used by the proponents, rather than to the methodology
underlying its success in accounting for the CMB
inhomogeneities. There are plenty of examples in fundamental
physics where the original motivations turned out not to be
justifiable but the idea was highly successful. I share the
view that the basic assumptions, listed above, are neither
`obvious' nor have they been justified from first principles.
However, the success of the inflationary paradigm with CMB
measurements is nonetheless impressive because one `gets much
more out than what one puts in'.

However, important open issues remain. First, there are issues whose origin lies in particle physics. Where does the inflaton come from? How does the potential arise? Is there a single inflaton or many? If many, what are the interactions between them? Since in the most commonly used scenarios the mass of the inflaton is very high, above $10^{12} {\rm Gev}$, the fact that we have not seen it at CERN does not mean it cannot exist. But in the inflationary scenario this is the only matter-field in the early universe and particles of the standard model are supposed to be created during `reheating' at the end of inflation when the inflaton is expected to roll back and forth around its minimum. In the modern viewpoint in cosmology, this is the substitute for the big-bang --simply a hot phase in the very early universe at which particles of the standard model are born, not the `absolute beginning' representing the earliest epoch of space-time in GR, represented by a space-like singularity. However, this putative reheating process is still poorly understood. What are the admissible interactions between the inflaton and the standard model particles which causes this decay? Does the decay produce the correct abundance of the standard model particles? These questions with origin in particle physics are wide open.

The second issue is the so-called `quantum to classical transition' referred to in the last assumption of standard inflation. As I explained there, one calculates the expectation values of perturbations and the two point function at the end of inflation and simply assumes that one can replace the actual quantum state of perturbations with a Gaussian statistical distribution of classical perturbations with the mean and variance, given by the quantum expectation value and the 2-point function. This strategy works very well as a practical devise for calculations. However, conceptual issues remain. Is there a \emph{physical} quantum to classical transition, induced by a  hitherto unknown process? Or, is the state still quantum mechanical but certain aspects of its behavior are well approximated by a classical distribution function? While this issue has drawn attention for well over a decade, in my view, it is only recently that our understanding has become sufficiently sharp: One can adopt the viewpoint that the standard description is viable because it is an \emph{excellent} approximation to the full quantum theory for the algebra of observables involving $n$-point functions of fields at the end of slow roll inflation (with $n \le 10$, say) \cite{ack}. These are the only observables that will be measured in the foreseeable future. While this result clarifies why the use of classical concepts works so well, the analysis \emph{simply assumes} that the initial quantum state of perturbations is of a certain type and does not say explain `why' or  `how' the universe happens to be in this state. I should also add that it does not provide a resolution of the standard `measurement problems' of quantum mechanics as is expected, in some quarters, of a fundamental cosmological theory.

The third set of issues has its origin in quantum gravity. In the standard inflationary scenario, one specifies initial conditions at the onset of inflation and then evolves the quantum perturbations. As a practical strategy, something like this is unavoidable within GR. Ideally one would like to specify the initial conditions at `the beginning', but one simply cannot do this in GR because the big bang is singular. Furthermore, since the curvature at the onset of
inflation is some $10^{-12}$ times the Planck scale,
by starting calculations at the onset, one bypasses the issue of the
correct Planck scale physics. But this is just an astute
stopgap measure. Given any candidate quantum gravity theory,
one can and \emph{has to} ask whether one can do better. Can
one meaningfully specify initial conditions in the Planck
regime? In a viable quantum gravity theory, this should be
possible because there would be no singularity and the Planck
scale physics would be well-controlled. If so, in the
systematic evolution from there, does a slow roll phase
compatible with the PLANCK mission data \cite{planck} arise
\emph{generically} or is an enormous fine tuning needed? One
could argue that it is acceptable to use fining tuning because,
after all, the initial state is very special. If so, can one
provide physical principles that select this special state? In
the standard inflationary scenario, if we evolve the modes of
interest back in time, they become trans-Planckian. Is there a
QFT on \emph{quantum} cosmological space-times needed to
adequately handle physics at that stage? Can one \emph{arrive
at} an observationally viable quantum state (at the onset of the WMAP slow roll) starting from natural initial conditions at the Planck scale?

In this article, I will not address the first two sets of issues.
Rather, the focus will be on the incompleteness related to the third
set, i.e., on quantum gravity. In section \ref{s3.2}, I will address a common question and in section \ref{s3.3} discuss a few concrete results to illustrate the current level of discourse in LQC.

\subsection{Why Pre-inflationary Dynamics Matters}
\label{s3.2}

\begin{figure}[]
 \begin{center}
  \includegraphics[width=2in,height=2.2in,angle=0]{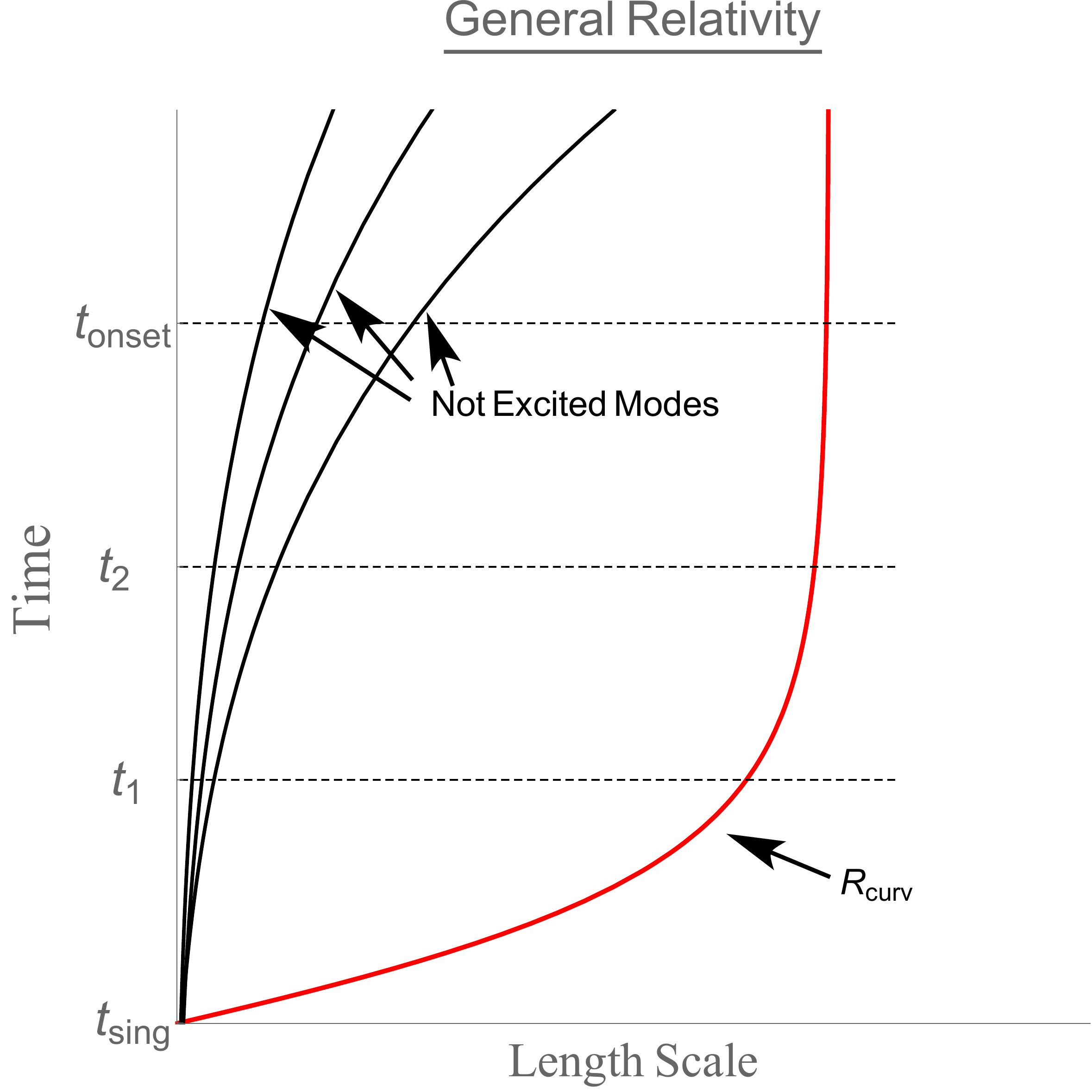}
\hskip2cm
\includegraphics[width=2.3in,height=2.2in,angle=0]{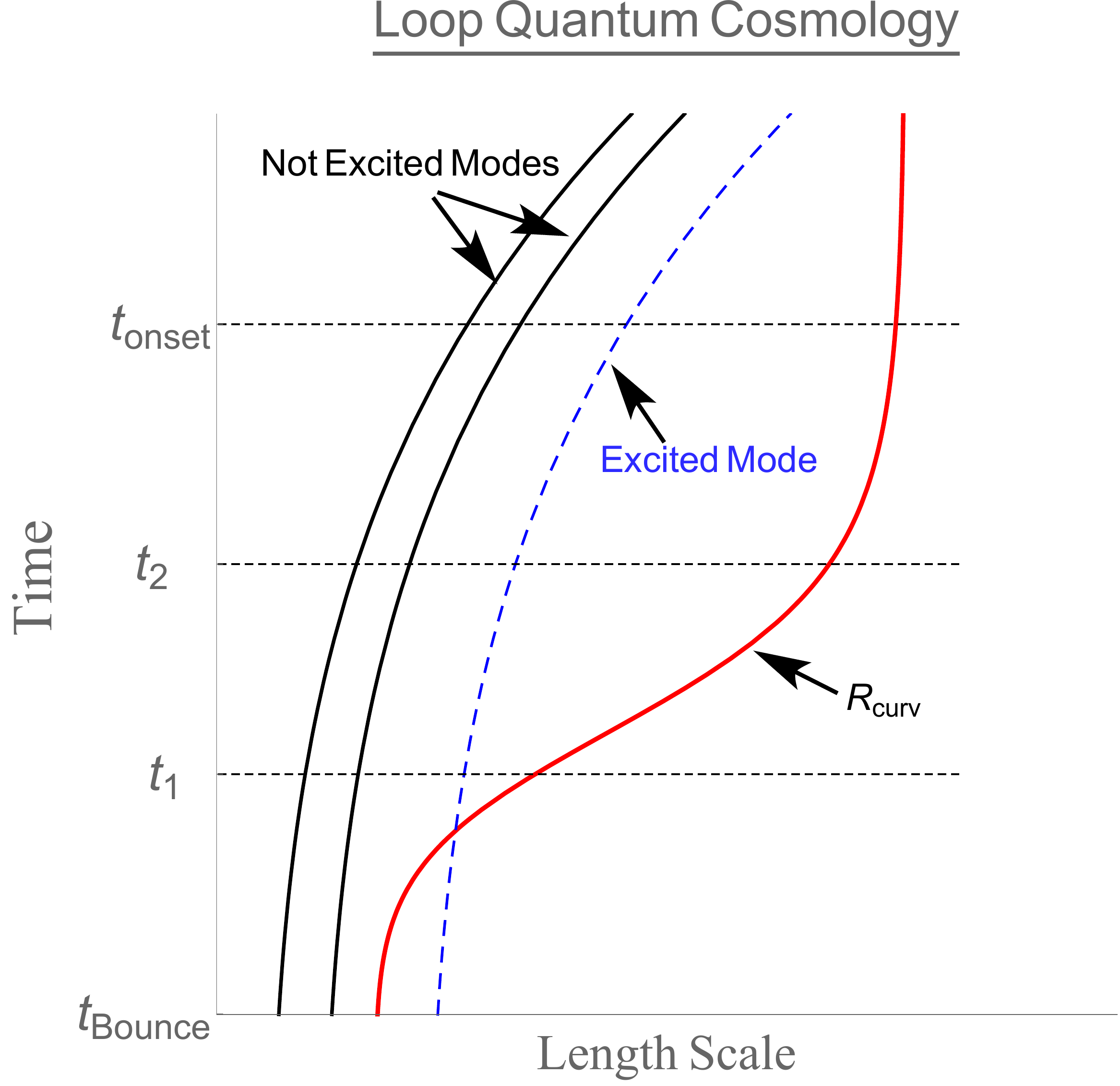}

\caption{\label{fig1} {\footnotesize{Schematic time evolution of the curvature radius ((red) solid line on the right in each panel) and of wave lengths of three modes seen in the CMB (three solid lines in the left panel and two solid lines and a (blue) dashed line in the right panel).\\
  \emph{Left Panel: GR.} The modes of interest have wave lengths less than the curvature radius all the way from the big bang
($t_{\rm sing}$) until after the onset of slow roll.\\
\emph{Right Panel: LQC.} The curvature radius is \emph{finite} at the big bounce ($t_{\rm Boun}$). Therefore, modes seen in the CMB can have wave lengths larger than the curvature radius in the very early universe. These modes exit the curvature radius in the Planck regime, experience curvature, and are therefore excited during the pre-inflationary evolution. They will not be in the BD vacuum at the onset of slow roll inflation. Figure Credit: B. Gupt.}}}
\end{center}
\end{figure}

At a fundamental level, of course, we need to understand pre-inflationary dynamics to determine whether the inflationary paradigm is part of a conceptually coherent framework encompassing the quantum gravity regime. But it is often claimed that, whatever the pre-inflationary dynamics may be, it can not change the observable predictions of the standard inflationary scenario. Indeed, this belief is invoked to justify why one starts the analysis just before the onset of the slow roll. The belief stems from the following argument, sketched in the left panel of Fig.~\ref{fig1}. If one evolves the modes that are seen in the CMB \emph{back} in time starting from the onset of slow roll, their physical wave lengths $\lambda_{\rm phy}$ continue to remain within the Hubble radius $1/H_{\rm GR}$ in the past. Therefore, one argues, they would not experience curvature and their dynamics would be trivial all the way from the big bang to the onset of inflation; because they are not `excited', all these modes would be in the BD vacuum at the onset of inflation. If pre-inflationary dynamics were indeed to have no observable effects, then we would have an interesting situation: While pre-inflationary dynamics is necessary for conceptual completeness, one would never be able to discriminate between various possibilities so long as they led to a standard slow roll phase.

However, the argument that pre-inflationary dynamics can not have observable effects is flawed on two accounts. First, if one examines the equation governing the evolution of these modes, one finds that what matters is the \emph{curvature radius} ${R}_{\rm curv} = \sqrt{6/\mathfrak{R}}$ determined by the Ricci scalar $\mathfrak{R}$, and not the Hubble radius. The two scales are equivalent only during slow roll on which much of the intuition in inflation is based. However, before the slow-roll phase, they are quite different from one another already in GR. Thus we should compare $\lambda_{\rm phy}$ with ${R}_{\rm curv}$ in the pre-slow-roll epoch. The second and \emph{much more important} point is that the pre-inflationary evolution should not be computed using GR, as is done in the argument given above. One has to use an appropriate quantum gravity theory since the two evolutions must be \emph{very} different in the Planck epoch if the quantum gravity effects are to resolve the big bang singularity. For, while in GR the ${R}_{\rm curv}$ goes to zero at the big bang since the Ricci scalar $\mathfrak{R}$ diverges there, in a satisfactory quantum gravity theory, 
${R}_{\rm curv}$ would remain non-zero. Therefore, \emph{there will always exist modes with  $\lambda_{\rm phy} \gtrsim {R}_{\rm curv}$} in the Planck epoch. These modes \emph{would be} excited during the phase in which the inequality holds. Therefore, the quantum state at the onset of the slow roll for these modes would be quite different from the BD vacuum. Our argument and hence the general conclusion essentially depends only on qualitative consideration that quantum effects do indeed resolve the singularity.%
\footnote{It also assumes that considerations of when the dynamics of modes is affected by the presence of scalar curvature carries over from quantum field theory in curved space-times to the Planck epoch. We will see in the next sub-section that this is indeed the case in LQC.} 

The key question then is whether these modes with $\lambda_{\rm phy} \gtrsim {R}_{\rm curv}$ in the Planck regime are seen in the CMB or whether they their physical wave length is larger than the observable universe at the CMB time. The answer to this question depends on the details of pre-inflationary dynamics. Since the physical wavelength $\lambda_{\rm phy}$ grows linearly with the scale factor, the question boils down to how many pre-inflationary e-folds there are.  Qualitatively, if this number $n_{\rm pre}$ is not too large, then the modes satisfying $\lambda_{\rm phy} \gtrsim {R}_{\rm curv}$ in the Planck regime \emph{will be} among the modes seen in the CMB. If $n_{\rm pre}$ is large, their physical wavelength in the CMB epoch will be too large to be observable. Thus, pre-inflationary dynamics of the specific quantum gravity theory under consideration will determine whether the pre-inflationary phase of evolution leaves an observable imprint on the CMB.

These considerations have opened a new window, with the potential to witness quantum gravity in action through observations of the very early universe. If the number of pre-inflationary e-folds is not too large, the predictions of that quantum gravity theory at the largest wavelengths will be quite different from that of standard inflation. 
In terms of the $C_{\ell}$ reported by observational missions, long wave lengths correspond to large angular scales, or small $\ell$. For low $\ell$, then, predictions a quantum gravity theory can differ from those of standard inflation. The difference could well be so large as to be incompatible with observations. In that case, the particular quantum gravity scenario would be ruled out. On the other hand, the differences could be more subtle: the new power spectrum for scalar modes could provide corrections to the standard inflationary predictions at small $\ell$ that provide a \emph{better fit} with observations. This is an exciting possibility because, interestingly, both WMAP and Planck missions have seen `anomalies' that cannot be accounted for by standard inflation precisely for $\ell\le 30$! Although these effects are only at a 2-3 $\sigma$ level, they raise a tantalizing possibility: they could well be the first glimpses of how Planck scale physics gets encoded in the CMB. There is also a potential that the same mechanism may lead to other departures from the standard predictions for observables that may be measured in future missions. Examples are: (i) suppression in the T-E and E-E power spectra at large angular scales (where T stands for temperature and E electric-type polarization) \cite{ab,ag2}, or, (ii) providing small but non-zero higher order T-T correlation functions leading to non-Gaussianities \cite{holman-tolley,agullo-parker,ganc,agullo-navarro-salas-parker,ia,ianm}. 

In LQC, the crucial number is $n_{\rm pre}$, the number of e-folds between the bounce and the onset of the slow roll inflation.
The right panel of Fig.~\ref{fig1} shows schematically the LQC pre-inflationary dynamics. (For the precise behavior obtained from numerical simulations, see Fig. 1 in \cite{aan3}.) As I will discuss in section \ref{s3.3}, there do exist quantum FLRW geometries on which wave lengths of some of the observable modes exceed the curvature radius during pre-inflationary dynamics, giving rise to departures from predictions of standard inflation. A highly non-trivial feature is that this phenomenon is a result of an unforeseen interlay between the ultraviolet and the infrared. Quantum geometry effects resolve the  singularity by taming the unruly behavior predicted by GR in the \emph{ultraviolet}. But this taming in turn opens a window for quantum gravity to leave its imprints on the largest wave length modes of perturbations, i.e., in the \emph{infrared}!

\subsection{LQC and Observations: Illustrative results}
\label{s3.3}

I will now present a few LQC results to illustrate the effects of pre-inflationary dynamics more concretely. 

Let us begin with the background dynamics. Systematic advances within LQC have extended the standard inflationary description all the way to the Planck scale. To do so, the classical FLRW geometries were replaced with a quantum FLRW geometry represented by a wave function $\Psi_{o}(v, \phi)$ which is sharply peaked around an effective trajectory of Eq. (\ref{effeq}). Detailed calculations were performed in the case of the simplest quadratic potential, $m^{2}\phi^{2}$. The main finding is that the evolution starting from the deep Planck regime does generically lead to the desired slow-roll phase of inflation \cite{svv,aads,ck,bl}. From the perspective of standard inflationary literature, this result is not surprising because, in presence of suitable potentials, inflationary trajectories have been shown to be attractors in GR. However, this result provides only a qualitative indication. The detailed calculations referred to above has sharpened the argument by showing that for the $m^{2}\phi^{2}$ potential there is a natural measure on the space of initial conditions \emph{at the bounce}. Using this measure, it has been shown that the probability that an initial data set at the bounce leads to a LQC effective trajectory that \emph{does not} enter the desired slow roll phase \emph{within the error-bars in the 9 year WMAP data} is less than $2.74\times 10^{-6}$ \cite{aads}.
 
In standard inflation, the linear cosmological perturbations are decomposed into scalar and tensor modes and, as explained in section \ref{s3.1}, the corresponding quantum fields --$\hat\Q$, the Mukahnov-Sasaki scalar mode, and $\hat{\T}_{I},\,\, I=1,2$,\, the two tensor modes-- propagate on the classical FLRW space-time. In LQC, one wishes to incorporate quantum geometry effects in the Planck regime. Two main directions are being pursued to carry out this task. One is based on the idea that the quantum corrections should be such that the resulting modifications of the GR constraints should still provide a closed Poisson algebra, free of `anomalies' \cite{bbcgk,mcbg,cmbg,cbgv,ed,bp,mlb,bm,lcbg,bgslb}. Because it relies on the Poisson algebra, it refers to effective equations \emph{satisfied by  perturbations} (not just the background quantum corrected geometry). This approach is described in Barrau's article in this special issue. 
By contrast, in the second approach, that I will describe here, both the background geometry and the perturbations are treated quantum mechanically: quantum fields $\hat\Q$ and $\hat{\T}_{I}$ now propagate on the \emph{quantum} FLRW geometry given by $\Psi_{0}$. As mentioned in section \ref{s1}, this approach has been developed in greater detail \cite{aan1,aan2,aan3,ia,ianm,bg1,bg2,fmo,cfmo,cmm} and is also well-suited to address some of the most important and most vexing conceptual problems \cite{ag2}. For a comparison between the two approaches, see, e.g., \cite{ab}.

Since $\Psi_{o}$ already incorporates quantum gravity effects that, in particular, tame the singularity, the second approach provides a systematic and natural avenue to face the so called `trans-Planckian issues' squarely. 
In particular, in solutions of interest, one finds that the observable universe evolved from a ball with surface area of $\sim 50\lp^{2}$ at the bounce. Now, the spectrum of the area operator in LQG is rather sophisticated: eigenvalues are not equally spaced but the level spacing decreases exponentially. Therefore nearby area eigenvalues are sub-Planckian. Consequently, LQG allows and can handle perturbations whose effective wave-length at the bounce are trans-Planckian.
The key question  is whether one can construct the required QFT on \emph{quantum} FLRW geometries $\Psi_{o}$. Since QFT in curved space-times already involves several conceptual and technical subtleties, at first the required extension to quantum background geometries appears to be intractable. However, an unforeseen simplification arises \cite{akl}. Let us suppose that the quantum state $\psi(\Q, \T;\, \phi)$ of perturbations is such that the back reaction of the quantum fields $\hat\Q,\, \hat{\T}_{I}$ on the background quantum geometry $\Psi_{o}(a,\phi)$ is negligible, i.e., that $\hat\Q,\, \hat{\T}_{I}$ can be treated as \emph{test fields}.%
\footnote{The test field approximation is also made in standard inflation. There, as well as in LQC, one justifies it by carrying out a self-consistency check that it does hold in the final solution. Note that one uses $\Q$ as the basic variable, rather than the curvature perturbation $\R$, because $\R$ is ill-defined if $\dot\phi$ vanishes and this does occur in the pre-inflationary phase. To obtain the scalar power spectrum, in LQC one calculates $\R$ from the Mukhanov-Sasaki variable $\Q$ \emph{at the end of inflation.}}
Then, if we start out with an initial state of the form $\Psi_{o}\otimes\psi$ at the bounce, the evolved state continues to have the same tensor-product form, where $\Psi_{o}$ evolves as though there was no perturbation but the evolution of $\psi$ depends on $\Psi_{o}$. Under this assumption, one can recast dynamics of $\psi$ in a way that is \emph{completely tractable} using existing techniques from QFT in curved space-times \cite{akl,aan1,aan3}. 

For a precise statement of this result, let us begin with the tensor perturbations $\T_{I}$. In the classical theory, these fields satisfy the wave equation $\Box \T_{I} =0$ on the background FLRW geometry. In LQC, one can show that the dynamics of the state $\psi(\T_{I};\,\phi)$ on the quantum FLRW geometry $\Psi_{o}$ is \emph{completely} equivalent to that of a state evolving on a certain FLRW metric $\tilde{g}_{ab}$ which is `dressed' by quantum corrections. Although  $\tilde{g}_{ab}$ is smooth, its coefficients depend on $\hbar$ and, more importantly, have to be constructed from the quantum FLRW geometry $\Psi_{o}(a,\phi)$ in a specific fashion. Detailed  calculations \cite{akl} reveal that $\tilde{g}_{ab}$ is given by:
\be \label{qcg} \tilde{g}_{ab} dx^a dx^b \equiv d\tilde{s}^2 =
\tilde{a}^{2} (-d\tilde{\eta}^{2}\, + \,  d{\x}^2 )\, .\ee
where
\be \label{qpara} 
\tilde{a}^4 = \f{\langle \hat{H}_o^{-\f{1}{2}}\,
\hat{a}^4(\phi)\, \hat{H}_o^{-\f{1}{2}}\rangle}{\langle
\hat{H}_o^{-1}\rangle};\quad \quad
d\tilde{\eta} = \langle \h{H}_{o}^{-1/2}\rangle\, (\langle \h{H}_{o}^{-1/2}\, \h{a}^{4}(\phi)\, \h{H}_{o}^{-1/2} \rangle)^{1/2}\,\, d\phi\,\,. \ee
Here, all operators and their expectation values refer to the Hilbert space of the background FLRW quantum geometry: the expectation values are taken in the state $\Psi_o$,\, $\h{H}_{o}$ is the `free' Hamiltonian in absence of the inflaton potential,\, and $\h{a}(\phi)$ is the (Heisenberg) scale factor operator \cite{akl,aan3}. To summarize, then, if the test field approximation holds, evolution of $\psi(\T_{I};\,\phi)$ on the quantum geometry $\Psi_{o}$ can be described using QFT of $\h{\T}_{I}$ propagating on the quantum corrected FLRW metric $\t{g}_{ab}$ via $\tilde\Box \h{\T}_{I}=0$ . 

Next, let us consider the scalar perturbation $\h\Q$. In QFT on classical FLRW space-times, it satisfies  $(\Box + \U/a^{2}) \h{\Q} =0$  where $\U$ is a potential constructed from the background FLRW solution. In the LQC dynamics, this $\U$ is also dressed, and is replaced by \cite{aan3}
\be \label{qpot} \t{\U}(\phi) = \f{\langle \h{H}_o^{-\f{1}{2}}\,
\h{a}^2(\phi)\, \h{\U}(\phi) \h{a}^2(\phi)\, \h{H}_o^{-\f{1}{2}}
\rangle}{\langle \hat{H}_o^{-\f{1}{2}}\, \hat{a}^4(\phi)\,
\hat{H}_o^{-\f{1}{2}}\rangle}\, . \ee
Thus, the evolution equation of the scalar mode $\h\Q$ is now given by $(\tilde\Box + \tilde\U)\, \h{Q} = 0$. Note that the scalar modes `experience' the same dressed metric $\t{g}_{ab}$ as the tensor modes. It is evident from (\ref{qpara}) and (\ref{qpot}) that the expressions of the dressed metric and the dressed potential could not have been guessed a priori. They resulted from explicit, detailed calculations \cite{akl,aan3}. They `know' not only the effective trajectory (of Eq. (\ref{effeq})) on which $\Psi_{o}$ is peaked but also certain fluctuations in $\Psi_{o}$. Finally note that the equivalence is \emph{exact} in the test field approximation; it does not involve any additional assumptions, e.g., on the wavelengths of the modes. This is rather striking.\\

\emph{Remark}: The following analogy is helpful in making the main result intuitively plausible. Consider light propagating in a material medium. Photons have a complicated interaction with the molecules of the medium. But so long as they do not significantly affect the material itself ---i.e., so long as the test field approximation holds for photons--- their propagation is well-described by just a few parameters such as the refractive index and birefringence which can be computed from the microscopic structure of the material. In our case, the quantum geometry encoded in $\Psi_{o}$ acts as the medium and quantum perturbations $\h{Q},\h{\T}_{I}$ are the analogs of photon field. So long as the test field approximation holds, the propagation is not sensitive to all the details  of the state $\Psi_{o}$. It can be described using just three quantities, $\t{a},\, \t\eta$ and $\t{\U}$ that are extracted from $\Psi_{o}$ via (\ref{qpara}) and (\ref{qpot}).
\\

This result provides a natural strategy to analyze the pre-inflationary dynamics in LQC by proceeding in the following steps. \emph{(1)} Select a state $\Psi_{o}$ of the quantum FLRW geometry that is sharply peaked on an effective trajectory, which in turn is determined by the value $\phi_{\B}$ of the inflaton at the bounce.  $\phi_{\B}$ turns out to be the \emph{new mathematical parameter} that determines the \emph{new physical parameter} $n_{\rm pre}$, the number  of e-foldings in the pre-inflationary phase, i.e., between the quantum bounce and the onset of the slow roll inflation. As explained briefly in section \ref{s3.2}, the number $n_{\rm pre}$, in turn determines the physical wavelengths of CMB modes for which LQC effects are important.  \emph{(2)} Calculate the dressed, effective metric $\t{g}_{ab}$ and potential $\t{\U}$ starting from $\Psi_{o}$. \emph{(3)} Select initial conditions for the quantum state $\psi$ of perturbations at the bounce.  \emph{(4)} Evolve this state using QFT on the dressed FLRW metric $\t{g}_{ab}$ and calculate the power spectrum and other correlation functions at the end of inflation. \emph{(5)} Check if the test field approximation holds throughout the `quantum geometry regime'. If not, one has to discard the solution. But if the approximation does hold, then $\Psi_{o}\otimes\psi$ would be a \emph{self-consistent solution}, representing an extension of standard inflation over the 11 orders of change in curvature that separate the onset of inflation from the Planck scale. \emph{(6)} Finally, check the \emph{physical} viability of the solution by comparing the predicted of power spectra and correlation functions with observations. 

By now, LQC has matured sufficiently to complete all these steps \cite{aan1,aan2,aan3,ia,ianm,bg1,bg2,fmo,cfmo,cmm}. Furthermore, thanks to the development of a systematic QFT on \emph{quantum} space-times, there is sufficient control on the dynamics of perturbations in the Planck regime to meet the `trans-Planckian issues' \cite{brandenberger} head-on. Consequently we now have a consistent theoretical framework to deal with cosmological perturbations on quantum FLRW space-times, starting from the bounce to the end of slow roll inflation.

Detailed calculations have been made for the $m^{2}\phi^{2}$ and the Starobinsky potentials. The final results, of course, depend on the choice of initial conditions for the background quantum geometry and for quantum perturbations --i.e., of states $\Psi_{o}$ and $\psi$. The primary emphasis has been on showing that \emph{there do exist} states for which:\,\,{\it (i)} there is a slow roll phase with desired characteristics%
\footnote{These include the number of slow roll e-foldings and, for each potential, the values of Hubble and slow roll parameters that are compatible with the observed amplitude and spectral index for scalar modes.}
\cite{svv,aads,ck,bl};\,\, {\it (ii)} The power spectrum agrees with standard inflation for $\ell \gtrsim 30$, and hence with observations \cite{aan1,aan2,aan3,ia,ianm,bg1,bg2};\,\, {\it (iii)} But there is power suppression at large angular scales, i.e., $\ell \lesssim 30$ so that the $\chi^{2}$-fit to the data is better than that for standard inflation \cite{ab,ag2,bg1,bg2,ia,ianm};\,\, {\it (iv)} non-Gaussianities are well within the observational limits \cite{ia,ianm}; and, \,\,{\it (v)} there is a hemispherical asymmetry seen in the PLANCK data \cite{ia,ianm}. 

In this sense, LQC has not only extended the standard inflationary scenario over the 11-12 orders of magnitude in density and curvature, but also provided better fits to the data at large angular scales. It is worth noting that in their article \emph{2015 results XVI: Isotropy and statistics of the CMB}, the Planck collaboration noted that the anomalies may be ``the visible traces of fundamental processes occurring in the early universe.'' This expectation is being borne out in the recent LQC investigations.

Note, however, that while these results are interesting and non-trivial, by themselves they are not hard tests of LQC; the $\sim\!3\sigma$ anomalies could well be explained in other ways. They could be just a statistical artifact due to `cosmic variance' at low $\ell$. It has also been proposed that the power suppression at large angular scales can be explained through the integrated Sachs-Wolf effect if one fine tunes the evolutionary history of the universe \cite{isw1,isw2}. If this were the mechanism, power suppression would not have anything to do with the physics of the primordial universe. However, in that case, there would not be suppression in the E-E (i.e., electric polarization) correlation functions at large angular scales. On the other hand, if the mechanism were primordial as in LQC, there \emph{would} be E-E power suppression \cite{ab,ag2}. The PLANCK collaboration is yet announce their results for T-E (temprature-electric polarization) or E-E (electric-electric) correlation functions for $\ell \lesssim 50$. Thus, if future observations show that the (T-E and) E-E correlation functions are suppressed relative to the prediction of standard inflation for $\ell < 30$, one would have a clear signal in favor of an LQC-type origin of these `anomalies'. Although there may well be other primordial mechanisms, such an observation would considerably increase confidence in the specific directions that are currently pursued in LQC. Irrespective of the outcome, from a conceptual standpoint it is interesting that LQC has advanced sufficiently to motivate a discourse at this level of
detail.

While the results I just listed were obtained within the broad LQC framework, they used different states both for the background and perturbations. Also the underlying mechanisms for various effects are somewhat different: while couplings between the very long and observable modes are key to the results on the hemispherical anisotropy \cite{ia,ianm}, as is common in papers on standard inflation that do not discuss non-Gaussianities, the mode-mode coupling is ignored in other papers \cite{aan1,aan2,aan3,ag2,bg1,bg2}. Secondly, as I  mentioned, the main emphasis has been on establishing the existence of the Heisenberg states $\Psi_{o}$ and $\psi$ with desired properties. However, the investigation of the issue of \emph{uniqueness}  has also been underway for some time.  The goal of this nearly completed program \cite{ag2} is to provide a theory of initial conditions. More precisely, the idea is to select: (i) a very narrow class of states states $\Psi_{o}$ representing the quantum FLRW geometry using a `natural' principle that relates the LQG area gap with the maximum radius of the observable universe; and, (ii) a  very narrow class of $\psi$ of perturbations by extending Penrose's  Weyl curvature hypothesis to \emph{quantum} cosmology. This procedure greatly narrows down the initial conditions, thereby making predictions of inflationary cosmology essentially unambiguous within LQC. Once this narrow class is selected, one can make further predictions for \emph{future observations}, e.g., on the T-E and E-E power spectrum at large angular scales. It is anticipated that the Planck mission will release their data on these correlation functions for $\ell \gtrsim 50$ in the near future. If the data does not bear out the prediction, then one would have to abandon the new proposal for choosing initial conditions. If on the other hand the data does bear out the prediction, one would have greater faith this new theory of initial conditions, developed within LQC. That would also be an indication that the large angular scale anomalies may indeed be a signature of Planck scale physics.

\subsection{summary}
\label{s3.4}

To arrive at a coherent extension of the inflationary scenario over the 11-12 orders of magnitude in curvature and density all the way to the Planck regime, researchers in LQC had to develop a new conceptual framework, introduce novel mathematical tools and carry out high
precision numerical simulations in the pre-inflationary phase. The issues they faced are diverse: The meaning of time in the Planck regime; the nature of quantum geometry in the cosmological context; QFT on \emph{quantum} cosmological space-times; renormalization and regularization of composite operators needed to compute stress energy and back reaction; high accuracy numerics; and, relation between theory and the PLANCK data. The final framework revealed an unforeseen interplay between the ultraviolet and the infrared which makes it feasible for quantum gravity effects in the deep Planck regime to leave imprints on the longest wave length modes one can observe in the CMB. The level of maturity of the subject is reflected in the fact that, given suitable Heisenberg states $\Psi_{o}$ of the background FLRW quantum geometry and $\psi$ of perturbations, not only can one calculate power spectrum at the end of inflation starting all the way from the deep Planck regime, but one can also make predictions for future observations: T-E and E-E power spectra and higher order T-T correlation functions. Furthermore, effort is under way to pin point physical principles that will provide the initial conditions, i.e., determine the Heisenberg states $\Psi_{o}$ and $\psi$ within a very narrow range, thereby tremendously boosting the predictive power of the theory. But, as emphasized in section \ref{s3.1}, LQC does not address any of the particle physics issues: The inflaton and its potential is an input. A long term possibility is that these inputs come from higher curvature corrections to Einstein equations in LQG 
\cite{ma} as in Starobinsky's original calculation

To conclude, let me emphasize that there was no a priori reason to anticipate either of the two main findings: the extension of standard inflation to the Planck regime and possible mechanisms behind the large angular scale anomalies found by the (WMAP and) PLANCK collaboration. A  priori it would not have been surprising if the pre-inflationary dynamics of LQC were such that the predicted power spectrum were observationally ruled out for the ÔnaturalÕ initial conditions at the bounce. Indeed, this could well occur in generic bouncing scenarios.

\section{Discussion and Outlook}
\label{s4}

Recall that, as in any quantum cosmology, LQC starts with a truncation of GR to its cosmological sector --FLRW metrics together with linear perturbations on them. In the 1970s, such truncations were regarded as the `royal road' to probe the fate of cosmological singularities in quantum gravity. However, now that this procedure has matured and led to some potentially observable signatures of Planck scale physics, it is appropriate to step back and examine it more carefully. 

Is this procedure of first reducing the classical theory to the physically  relevant symmetric sector and then passing to quantum theory justified? Or, does one miss out some key features that would alter the findings in a qualitative fashion? There is an outstanding example in the history of quantum mechanics that provides interesting insights on this issue: the Dirac theory of hydrogen atom. Here one considers the proton-electron system, truncates the Maxwell theory to its \emph{spherical symmetric sector} and then carries out quantization. From the perspective of full quantum electrodynamics (QED) the strategy seems to introduce  \emph{drastic} oversimplifications since it banishes all the photons right from the start! Indeed, conceptually, the truncated theory does ignore most of the rich possibilities one can envisage in full QED. And yet the Dirac theory provides an excellent description of the hydrogen atom. Indeed, to see its limitations, one has to carry out \emph{very} accurate measurements and carefully examine the hyperfine structure due to QED effects such as the Lamb shift! Returning to quantum cosmology, this example illustrates why it is not unreasonable to expect the symmetry reduction strategy would be successful. In LQC, quantum theory is \emph{not} constructed in a manner that is engineered just to the symmetry reduced model under consideration. The procedure pays due attention to quantum geometry that emerged from \emph{full} LQG.%
\footnote{More precisely, in kinematics of full LQG, one starts with the  diffeomorphism covariant ``holonomy-flux quantum algebra'', finds the unique regular, diffeomorphism invariant vacuum state on it \cite{lost,cf} and generates other quantum states by the action of the algebra on the vacuum \cite{alrev,crbook,ttbook}. It is because the procedure is background independent that the geometric observables have discrete eigenvalues. In LQC one starts with the holonomy-flux algebra adapted to homogeneous isotropic fields, shows that it admits a unique regular state that is invariant under the (residual) diffeomorphisms and generates other states by the action of the algebra on this vacuum \cite{aamc}. Thus, one follows exactly the same procedure as that used in full LQG. Several results obtained over the past 2 years are now paving the way to bridge LQG and LQC in detail \cite{ilqg}.}
Therefore, for the few observables that are needed to describe the very early universe, the varied and rich possibilities of full quantum gravity could just refine, rather than alter, the LQC description.

Consider in particular the issue of singularity resolution. As discussed in section \ref{s2}, LQC provides a natural singularity resolution not only in the FLRW models but also in the anisotropic Bianchi models. Now, in classical GR, there is a long standing conjecture due to Belinskii Khalatnikov Lifshitz (BKL) that posits that, as one approaches a generic space-like singularity in GR, the local evolution is well approximated by the Bianchi I, II and IX models \cite{bkl,ahs1}. Therefore the fate of singularities in Bianchi models is of special interest. A common concern has been that even if the big bang is replaced by a big bounce in the isotropic case, typically this singularity resolution would not survive in Bianchi models (primarily because the anisotropic shear terms diverge as $1/a^6$ where $a$ is the scale factor). Indeed, this is the case in several approaches to quantum cosmology \cite{asrev,ps3}. In LQC, by contrast, results on singularity resolution do extend to these Bianchi models \cite{awe2,madrid-bianchi,awe3,we,gs,swe,ps2}. Furthermore, if one traces the Hamiltonian constraint operator of the Bianchi I model over anisotropies, one is led precisely to the FLRW Hamiltonian constraint operator, bringing out robustness of the scheme \cite{awe2}. This is a small but concrete indication that the mechanism that has been uncovered in the LQC of FLRW models may not be an artifact of symmetry reduction, but captures an essential feature of full LQG (see also Alesci's contribution to Ref. \cite{ilqg}). Indeed, in view of the BKL conjecture, there are concrete reasons to hope that quantum geometry effects will cure all space-like, strong singularities of classical GR \cite{asrev,ps3,ps1}.   

What about the truncation used in treating cosmological scalar and tensor perturbations $\hat{\mathcal{R}}$ and $\hat{\mathcal{T}}$? LQC restricts itself to states of the type $\Psi_o\otimes\psi$ where $\Psi_o$ is a state of the quantum FLRW geometry and $\psi$ is the state of linear quantum perturbations $\hat{\mathcal{R}},\,\hat{\mathcal{T}}$. Full LQG will of course admit states in which there are huge quantum fluctuations in the Planck regime whose physics cannot be captured by states of this type. But there is a tendency to go overboard and assume that \emph{all} states of the full quantum gravity will have huge fluctuations in the Planck regime. At the theoretical level, LQC has provided concrete evidence that this need not be the case: there do exist states of the type $\Psi_o\otimes\psi$ for which truncation is \emph{self consistent}. These states lead to an unforeseen, tame behavior in which $\hat{\mathcal{R}},\,\hat{\mathcal{T}}$ evolve as linear perturbations on a background quantum geometry $\Psi_o$. The non-triviality lies in the fact that these self consistent, truncated solutions lead to the power spectrum and spectral index that are consistent with observations. Thus, the situation is similar to that in the standard $\Lambda$CDM\index{$\Lambda$CDM} model where it suffices to restrict oneself to the simplest cosmological solutions. A priori, just on theoretical grounds one would not have expected the early universe to be as simple as it has turned out to be. Agreement between observations and self-consistent LQC solutions suggests that this simplicity extends also to the Planck regime. 

This does \emph{not} mean that full LQG will not admit states which are very different from the simple $\Psi_{o}\otimes\psi$. Indeed in the classical regime full, non-linear GR does admit cosmological solutions which are much too complicated to be approximated by FLRW geometries together with first order perturbations even in the early era. It is just that observations show that these solutions are not realized in the physical universe we inhabit. On the quantum side, it is clear that the structure of quantum gravity will be much more complicated than that revealed by symmetry reduced models. Surely full quantum gravity admits states with wild quantum fluctuations. The issue is rather whether these complications are relevant for specific \emph{physical} problem of describing the large scale structure of \emph{our} universe. The detailed LQC results suggest that, if one restricts oneself to the few quantities that are relevant to observations of the very early universe, the answer is likely to be in the negative.\\

I will conclude with two examples that further elucidate this point, one from cosmology and another from black holes. 

Recall that cosmological singularities were first encountered in highly symmetric solutions of Einstein's equations. Serious research groups in mathematical GR believed that the models were too simple --just `toy' models' in light of the complexity of the full theory-- to teach us anything useful. Indeed, there was a wide spread belief that generic solutions to full Einstein's equations, without any symmetries, would be \emph{singularity-free}  because the full complexity of the GR would reveal qualitatively new features that are lost in the symmetry reduction procedure. Powerful singularity theorems of Penrose, Hawking, Geroch and others showed that this attitude was misguided. Interestingly, we now have a reciprocal situation at the quantum level. Symmetry reductions of LQG have shown that singularities \emph{are} resolved due to the underlying \emph{quantum} geometry. Rather than regarding these only as results from `toy models,' it would be more fruitful to consider them as providing guidance for the full theory. In classical GR it is the rather simple-looking Raychaudhuri equation that opened the door to the general singularity theorems. Is there the figurative analog of the Raychaudhuri equation in full LQG that would reveal that \emph{all} space-like strong curvature singularities of GR are resolved in LQG? 

The second quantum gravity area in which we are faced with concrete physical issues concerns black holes. In contrast with cosmology, here we are not guided by direct observations. Rather, issues are brought into sharp focus by theoretical puzzles: the `obvious' avenues and expectations appear to lead to conclusions that violate one or more of our cherished principles/beliefs. Since the analysis of the full problem is technically very difficult, again one focuses on truncations that appear to capture all the important conceptual issues, without waiting for a complete quantum gravity theory.

Specifically, in recent years restriction to spherically symmetric space-times has led to several advances. First, a further truncations to quantum field theory on classical black hole space-times led to Hawking's seminal discovery of black hole evaporation. Continued research has often involved simplified models. This work has led to a number of unforeseen insights. For example, inclusion of back-reaction through a mean field approximation has revealed unforeseen features such as a universal behavior at the end of the semi-classical process and a correction to the Bondi-energy formula that had been used \cite{atv,apr2}. Another line of investigation has led to concrete results on relations between entanglement entropy and energy radiated across null infinity \cite{bs,bds}. Finally, quantum geometry of spherical black holes has now been analyzed in detail \cite{gp1}. Furthermore, following the use of quantum fields on quantum cosmological space-times discussed in section \ref{s3.3}, the evaporation process has been studied using quantum fields on quantum black hole geometries \cite{gp2,gop}. This investigation has opened the door to understanding the evaporation process in the LQG framework of full quantum gravity.\\ 

Finally, I wish to emphasize that focus on symmetry reduced does not diminish the importance of the task of constructing a consistent and viable mathematical framework for the full theory. This construction is of paramount importance not only for the obvious reason of the need for a coherent conceptual framework but also because, once it becomes available, it will inevitably lead us to new questions and probe new phenomena that we cannot even dream of today. But it would be incorrect to ignore the \emph{physical} insights we are gaining from truncations to cosmology and black holes. They are beacons, pointing out valuable opportunities and fertile directions.

\section*{Acknowledgments}

Over the years, I have profited by discussions on the subject of this review with a large number of colleagues. I would especially like to thank  Ivan Agullo, Aurelien Barrau, Alejandro Corichi, Brajesh Gupt, Wojciech Kaminski, Jerzy Lewandowski, William Nelson, Thomasz Pawlowski, Parampreet Singh and Edward Wilson-Ewing. Because this is an overview, there is inevitable overlap with several research articles as well as reviews I have written on this subject, in particular, \cite{asrev,prague,arr,ab}. This work was supported in part by the NSF grant PHY-1505411 and by the Eberly endowment funds at Penn State.


\begin{thebibliography}{99}
\bibitem{ps2}P.~Singh, Is classically flat Kasner spacetime flat in quantum gravity?  \texttt{arXiv1604.03828}
\end{thebibliography}
\end{document}